\documentclass[endfloats,aps,prb,twocolumn,groupedaddress]{revtex4}

\usepackage{graphicx}

\begin{document}


\title{Magnetic and Electronic Phase Diagram and Superconductivity in the Organic Superconductors $\kappa$-(ET)$_{2}$$X$}


\author{T. Sasaki}
\author{N. Yoneyama}
\author{A. Matsuyama}
\author{N. Kobayashi}
\affiliation{Institute for Materials Research, Tohoku University, Sendai 980-8577, Japan}



\date{\today}

\begin{abstract}
The magnetic susceptibility of the organic superconductors $\kappa$-(h8 or d8-ET)$_{2}$$X$, $X =$ Cu(NCS)$_{2}$ and Cu[N(CN)$_{2}$]Br has been studied.  
A metallic phase below $T^{*} =$ 37 $\sim$ 38~K for $X =$ Cu[N(CN)$_{2}$]Br and 46 $\sim$ 50~K for $X =$ Cu(NCS)$_{2}$ has an anisotropic temperature dependence of the susceptibility and the charge transport.  
Partial charge-density-wave or charge fluctuation is expected to coexist with the metallic phase instead of the large antiferromagnetic fluctuation above $T^{*}$.  
The phase diagram and the superconductivity of $\kappa$-(ET)$_{2}$$X$ are discussed in connection with this phase.
\end{abstract}

\pacs{74.70.Kn, 74.25.Dw, 74.25.Ha, 74.62.Fj}

\maketitle


Organic charge transfer salts based on the donor molecule bis(ethylenedithio)-tetrathiafulvalene, abbreviated BEDT-TTF or ET, are characterized by their quasi-two dimensional (Q2D) electronic properties. \cite{Singleton1}  
Among them, the $\kappa$-type organic superconductors, $\kappa$-(ET)$_{2}$$X$ with $X =$ Cu(NCS)$_{2}$ and Cu[N(CN)$_{2}$]$Y$ ($Y =$Br and Cl) have attracted considerable attention from the point of view of the strong electron correlation effect and the superconductivity. \cite{Kanoda1,McKenzie1}  
The phase diagram shows that the antiferromagnetic (AF) ordered state is in contact with the superconducting phase and normal state properties are quite different from the conventional metals. \cite{Kanoda1,McKenzie1}  
Since the similarity of the phase diagram and some unusual properties in the normal state imply analogies to the high-$T_{c}$ cuprates with carrier doping playing the role of pressure in organics, \cite{McKenzie1} the AF spin-fluctuation was expected to be the origin of the superconductivity. \cite{Kino,Kondo,Kuroki,Schmalian,Louati}
In fact, the spin-lattice-relaxation rate $(T_{1}T)^{-1}$ of $^{13}$C-NMR in such superconducting salts shows an enhancement below 100 $\sim$ 150~K and takes a cusp around $T^{*} \simeq$ 50~K. \cite{Mayaffre1,Kawamoto1,Soto1}  
The enhancement and the anomaly at $T^{*}$ have been interpreted as AF spin fluctuation and a pseudogap formation.  
Recently large softening of the ultrasound modes and the pronounced minima at $T^{*} = 37 \sim$ 38~K in $X =$ Cu[N(CN)$_{2}$]Br salt and 46 $\sim$ 50~K in $X =$Cu(NCS)$_{2}$ salt have been observed. \cite{Frikach,Shimizu} 
The softening was attributed to the coupling between acoustic phonons and AF fluctuations .  
Then the importance of AF spin fluctuation is generally agreed above $T^{*}$, but the rapid restoration of the softening and the cusp behavior of $(T_{1}T)^{-1}$ are not always understood as occurring pseudogap formation at $T^{*}$.  
Very recent thermal expansion measurements reveal that not a crossover such as pseudogap formation but a second-order phase transition takes place at $T^{*}$. \cite{Mueller1}  
The important point is what the phase between $T^{*}$ and $T_{c}$ is, because the superconductivity should be considered on the basis of this phase.  

Concerning the mechanism and the symmetry of the superconducting order parameter, AF spin-fluctuation induced superconductivity with the pairing symmetry of $d_{x^{2}-y^{2}}$ is theoretically proposed. \cite{symmetry} 
Although the experimental investigations have been continued intensively, the situation is not settled \cite{Lang1}: the reported results suggest conventional BCS-like behavior or unconventional $d$-wave state with line node gap.  
The recent gap direction sensitive experiments as STM \cite{Arai} and thermal conductivity \cite{Izawa} predict the line node gap rotated 45$^{\circ}$ relative to the $b$ and $c$-axes ($d_{xy}$ symmetry), while the millimeter-wave transmission experiment suggests nodes along the $b$ and $c$-axes ($d_{x^{2}-y^{2}}$ symmetry). \cite{Schrama}  
(An alternative interpretation was proposed for the latter result. \cite{comments})
The former result of the $d_{xy}$ symmetry, which is favored for charge fluctuations while $d_{x^{2}-y^{2}}$ is favored for AF fluctuations, \cite{Scalapino} is inconsistent with the AF fluctuation scenario.  
This inconsistency may be closely related to the phase transition at $T^{*}$ above $T_{c}$.  

In this paper, the systematic measurements of the magnetic susceptibility of $\kappa$-(h8 or d8-ET)$_{2}$$X$ with X=Cu(NCS)$_{2}$ and Cu[N(CN)$_{2}$]Br are reported.  
We focus on the anisotropic behavior below $T^{*}$ in order to know the nature of the phase.  
On the basis of the phase diagram proposed in this study, we discuss the possible nature of the superconductivity of $\kappa$-(ET)$_{2}$$X$.  

The hydrogenated and deuterated ET donor molecule were used for the electrocrystallization.  
These two type of crystals are denoted as $\kappa$-(h8 or d8-ET)$_{2}$$X$.  
The magnetic susceptibility measurements were performed using a SQUID magnetometer (Quantum Design, MPMS-5) in $H =$ 5~T, excepting the Meissner effect measurements in 0.5~mT.  
The data are corrected for the demagnetization factor of each sample shape.

Figure~1(a) shows the temperature dependence of the magnetic susceptibility of a single crystal (3.0 mg) of $\kappa$-(h8-ET)$_{2}$Cu(NCS)$_{2}$ in the field parallel to the three crystal axes; the $a^{*}$ (perpendicular to the Q2D plane), $b$ and $c$-axes (in the Q2D plane).  
The inset depicts the observed susceptibility $\chi$ after subtracting the core diamagnetic contribution $\chi_{\rm core}$ ($-4.7 \times 10^{-4}$ emu/mole) which is regarded as isotropic and constant with temperature in this panel.  
The temperature dependence is very similar to the previous report \cite{Toyota}: temperature independent behavior above 100~K and gradual decrease below.  
Added to these, slight temperature independent anisotropy is observed in the present high sensitive measurements. 
This anisotropy may come from anisotropic contribution of $\chi_{\rm core}$, which is expected from a stack of the planer ET molecules but is not well known at present.  
In the main panel the susceptibility $\chi_{\rm spin}$ is plotted by shifting the data along the $b$ and $c$-axes to the value along the $a^{*}$-axis by small constants of $+0.19$ and $+0.20 \times 10^{-4}$ emu/mole, respectively.  
The gradual decrease of $\chi_{\rm spin}$ follows an activation type temperature dependence, \cite{Kataev,Gruner} $\chi_{\rm spin} \propto (1/T)\exp(-\Delta/k_{B}T) + \chi_{0}$, which is drawn by the solid line where $\Delta =$ 101~K and $\chi_{0} = 3.26 \times 10^{-4}$ emu/mole.  
This activating temperature dependence is in agreement with a scenario of AF or spin-density-wave (SDW) fluctuation above $T^{*}$.  
The existence of the constant term $\chi_{0}$ may imply that a residual metallic contribution mainly comes from the closed part of Fermi surface (FS).
Ascending deviation from the activating temperature dependence starts at $T^{*} \simeq$ 45~K. \cite{errors}  
This change at $T^{*}$ means that the system becomes more metallic where AF spin fluctuation tends to be suppressed.  
This corresponds to the suppression of $(T_{1}T)^{-1}$ in $^{13}$C-NMR. \cite{Mayaffre1,Kawamoto1,Soto1}  
Besides these $\chi_{\rm spin}$ shows weak anisotropic behaviour ($\chi_{b} > \chi_{c} \simeq \chi_{a^{*}}$) below $T^{*}$.  
Such anisotropic temperature dependence is more clearly seen in $\kappa$-(h8-ET)$_{2}$Cu[N(CN)$_{2}$]Br.  
Figure~1(b) shows the magnetic susceptibility on aligned three single crystals (total 2.7 mg) to the $b$-axis.  
The magnetic field is applied parallel to the $b$-axis (perpendicular to the Q2D plane) and the $a$-$c$ plane (parallel to the Q2D plane).  
The inset depicts $\chi$ after subtracting $\chi_{\rm core}$ ($-4.8 \times 10^{-4}$ emu/mole).  
Overall features are very similar to those in $\kappa$-(h8-ET)$_{2}$Cu(NCS)$_{2}$.
In the main panel, $\chi$ along the $a$-$c$ plane is plotted after shifting by a constant of $-0.37 \times 10^{-4}$ emu/mole.  
The solid line is a fitted result to $\chi_{\rm spin}$ above 50~K by the same activating type temperature dependence with $\Delta =$ 102~K and $\chi_{0} = 2.60 \times 10^{-4}$ emu/mole.    
Anisotropic behavior below $T^{*} \simeq$ 35~K is more pronounced than that in $\kappa$-(h8-ET)$_{2}$Cu(NCS)$_{2}$. 
Magnitude of the anisotropy at 20~K is roughly estimated to be 1.5\% for $\kappa$-(h8-ET)$_{2}$Cu(NCS)$_{2}$ and 7\% for $\kappa$-(h8-ET)$_{2}$Cu[N(CN)$_{2}$]Br. 
The difference of the anisotropy may be related to the same tendency with the size of the softening in the ultrasound velocity \cite{Frikach} and the volume expansion coefficient at $T^{*}$. \cite{Mueller1}  

We now move on to the relation between the superconductivity and the susceptibility behavior.  
It has been examined that the superconductivity of $\kappa$-(d8-ET)$_{2}$Cu[N(CN)$_{2}$]Br can be controlled by the cooling speed. \cite{Kawamoto2,Nakazawa1}  
Figure~2 demonstrates the different susceptibility behavior of the slow cooled (0.2~K/min) and the quenched (100~K/min) sample of not-aligned several crystals (total 7.9~mg).  
Small anomaly around 45~K is not intrinsic but may be due to the magnetic transition of the residual solid oxygen. 
The change of the superconductivity with the cooling is seen in the inset.  
The sample after the quenched process shows almost no superconductivity with less than 0.1\% of the Meissner volume, while about 10\% of the volume becomes superconducting at 5~K after the slow cooled process.  
In the main panel, the difference of $\chi_{\rm spin}$ for the two cooling process appears below about 30~K, where $\chi_{\rm spin}$ in the slow cooling is larger than that in the quenched process.  
It is also noted that these $\chi_{\rm spin}$'s, especially the quenched one, are smaller than the value expected from the activation type behavior with $\Delta =$ 102~K and $\chi_{0} = 2.90 \times 10^{-4}$ emu/mole.  
The quenched $\chi_{\rm spin}$ can be explained as follows.  
The AF fluctuation becomes enhanced continuously with decreasing temperature and the AF static order appears at $T_{\rm N} \simeq$ 17~K where $\chi_{\rm spin}$ takes a minimum.  
Then no transition appears in the trace of the temperature dependence and no $T^{*}$ exists in the non-superconducting sample.  
This is also consistent with the thermal expansion measurements in non-superconducting salts which show no corresponding anomaly observed in the superconducting sample at $T^{*}$. \cite{Mueller1}  
The sample in the slow cooled process shows an intermediate behavior.  
Both weak AF order at $T_{N} \simeq$ 17~K and weak superconductivity at $T_{c} \simeq$ 11.5~K are expected to exist inhomogeneously in the sample.  
The difference of $\chi_{\rm spin}$'s below 30 $\sim$ 35~K can be regarded to the paramagnetic contribution from a superconducting part of the slow cooled $\chi_{\rm spin}$.  
Therefore $T^{*}$ of the slow cooled $\kappa$-(d8-ET)$_{2}$Cu[N(CN)$_{2}$]Br, which has a superconducting volume fraction, is naturally expected to be in the range of 30 $\sim$ 35~K.
It is in agreement with the previous observation of small hump of $(T_{1}T)^{-1}$ around 30~K in the slow cooled sample. \cite{Kawamoto2}  

Let us consider the anisotropic behavior below $T^{*}$ from the electronic conductivity point of view.  
Figure~3 shows the temperature dependence of the resistance along the $b$ and $c$-axes of $\kappa$-(h8-ET)$_{2}$Cu(NCS)$_{2}$.  
The measurements were performed by means of the standard {\it dc} four-probe method using perpendicularly arranged two sets of four electrical contacts on one crystal.  
The characteristic feature of the resistance is very similar to the previous reports: a large hump around 100~K and change of the temperature dependence around 50~K.  
In addition, the present results clearly demonstrate that the anisotropic temperature dependence of the charge transport appears below about 50~K corresponding to $T^{*}$.  
The bottom right inset shows the resistance ratio $R_{b}/R_{c}$ normalized at 273~K.  
A steep increase starts around $T^{*} \sim$ 50~K, which suggests that the charge transport along the $b$-axis becomes resistive than that along the $c$-axis.  
In contrast to the sharp increase at $T^{*}$, the anisotropy of the charge transport does not show noticeable change through the AF fluctuation region.

In view of these experimental results, let us then consider the phase diagram.  
Figure~4 summarizes both the present results and the recent precise report of the pressure effect on $\kappa$-(h8-ET)$_{2}$Cu[N(CN)$_{2}$]Cl by Lefebvre {\it et al}. \cite{Lefebvre}
A relative pressure is taken as the horizontal axis and the position of the various salts at ambient pressure is indicated by the dotted arrows. \cite{Kanoda1}   
The solid lines of $T_{c}$, $T_{\rm N}$ and $T^{*}$ refer to the results of the pressure dependence studies. \cite{Frikach,Lefebvre,Schirber,Caulfield}  
It is interestingly worth mentioning that the $T^{*}$ line seems to be terminated at a critical point (220~bar, 32.5~K) of the metal-insulator transition in $\kappa$-(h8-ET)$_{2}$Cu[N(CN)$_{2}$]Cl. \cite{Lefebvre}  
The main finding of the present study is that the region between $T^{*}$ and $T_{c}$ is a phase (PM/DWF) with more metallic nature but anisotropic $\chi_{\rm spin}$ and charge transport in contrast to the metallic phase with large AF spin fluctuation (PM/AFF).  
The anisotropic behavior of $\chi_{\rm spin}$ and charge transport in PM/DWF may suggest that the phase is accompanied with SDW and/or charge-density-wave (CDW) instability on the open part of FS.  
In the present case of $\kappa$-(h8-ET)$_{2}$Cu(NCS)$_{2}$, the $b$-direction is expected to be a good nesting direction for the quasi-one dimensional(Q1D) part of FS.  
The anisotropy of the charge transport can be understood in this picture: the nesting gap on the Q1D band has an influence mainly on the $b$-axis conductivity being resistive. 
Such density-wave transition and the gap formation is consistent with STM results of a broad gap structure persisting above $T_{c}$ and up to $T^{*}$. \cite{Arai2}
The SDW scenario is, however, difficult to explain that no broadening of the line width in $^{13}$C-NMR spectra is observed at $T^{*}$. \cite{Kawamoto1,Soto1,Mayaffre2}
Then we propose that the PM/DWF phase is a metallic state with CDW or charge fluctuation.  
The anisotropy of $\chi_{\rm spin}$ may be explained by a CDW model, which predicts the susceptibility is most paramagnetic when the magnetic field and the CDW vector are parallel. \cite{Boriack}  
It is consistent with the observation of the largest $\chi_{\rm spin}$ along the $b$-axis.  
On the other hand, the non-superconducting salts located below the critical pressure have only one second-order phase transition from the paramagnetic insulating phase with the large AF fluctuation (PI/AFF) to the insulating AF static order phase (AFI) at $T_{\rm N}$. \cite{Miyagawa}  
Gradual change from the paramagnetic and non-metallic phase (PNM) at high temperature to PM/AFF and PI/AFF is expected to be a crossover at which the AF fluctuation starts growing with a spin gap (pseudogap) formation. \cite{Mayaffre1,Kawamoto1,Soto1}
It is noted that the superconducting phase (SC) is realized by the second-order transition from PM/DWF, while it is only stabilized by the first-order transition from AFI as a function of pressure. \cite{Lefebvre}  
Then it seems reasonable to suppose that SC of $\kappa$-(ET)$_{2}$$X$ should be considered with the weak coupling scenario from the PM/DWF side, not with the strong coupling scenario from the AFI side.  
The superconductivity based on the CDW and charge fluctuation has been suggested to have the $d_{xy}$ symmetry. \cite{Scalapino,Merino1}  
Recent STM \cite{Arai} and the thermal conductivity \cite{Izawa} results on the gap symmetry may have close relation with this issue.  

In summary, the magnetic susceptibility of $\kappa$-(ET)$_{2}$$X$ was studied to examine the phase diagram.  
The PM/DWF phase with CDW or charge fluctuation is proposed to exist only in the superconducting salts.  
The charge fluctuation scenario is likely to be realized in this class of the organic superconductors.  

The authors thank M.~Lang, Y.~Matsuda, K.~Miyagawa, M.~Mori, J.~M\"uller and T.~Tohyama for stimulating discussions.  
We are grateful to T. Fukase for his encouragement.  
This work was partially supported by a Grant-in-Aid for Scientific Research from the Ministry of Education, Science, Sports and Culture of Japan.

%

\bibliography{basename of .bib file}

\begin{thebibliography}{}
\bibitem{symmetry} In this paper, the coordinates of the order parameters, for example $d_{x^{2}-y^{2}}$ and $d_{xy}$, are taken in the extended magnetic Brillouin zone in the similar style of the High-$T_{c}$ cuprates. \cite{Kino,Kondo,Kuroki,Schmalian}  See the inset of Fig.~3.
\bibitem{errors} Precise determination of $T^{*}$, with the error bars of $\pm$~5~K, is difficult from the susceptibility data because of gradual growth of the anisotropy and small magnetization observed with the accuracy of about 1~\% in $\kappa$-(h8-ET)$_{2}$$X$ and 0.5~\% in $\kappa$-(d8-ET)$_{2}$Cu[N(CN)]$_{2}$Br.
\end{thebibliography}

\begin{figure}
\includegraphics[viewport=2cm 3cm 20cm 26cm,clip,width=0.9\linewidth]{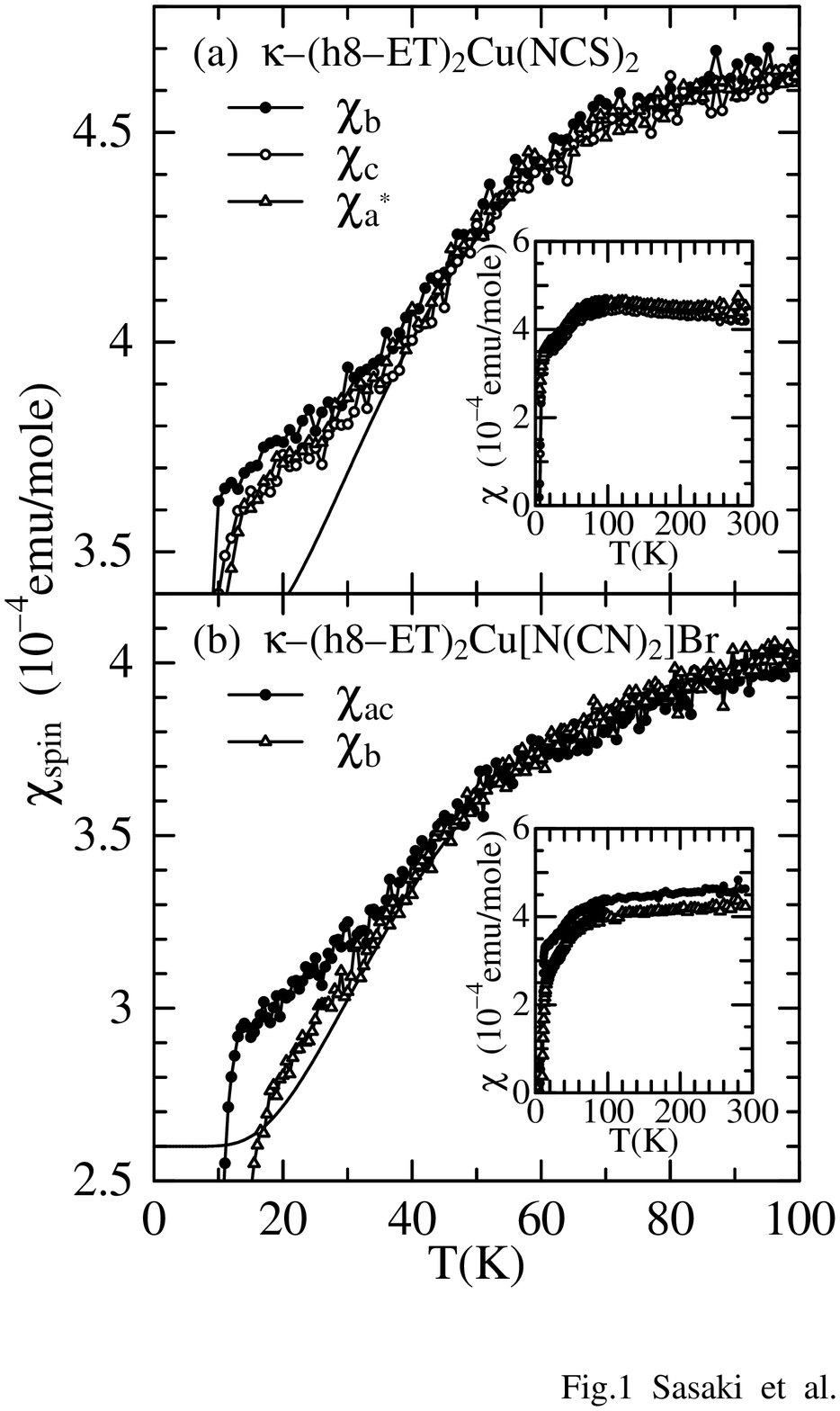}
\caption{Temperature dependence of the magnetic susceptibility of (a) $\kappa$-(h8-ET)$_{2}$Cu(NCS)$_{2}$ and (b) $\kappa$-(h8-ET)$_{2}$Cu[N(CN)$_{2}$]Br in 5~T.  The solid curves show an activating type temperature dependence fitted to the data above 50~K.}
\end{figure}

\begin{figure}
\includegraphics[viewport=2cm 7.5cm 20cm 23cm,clip,width=0.9\linewidth]{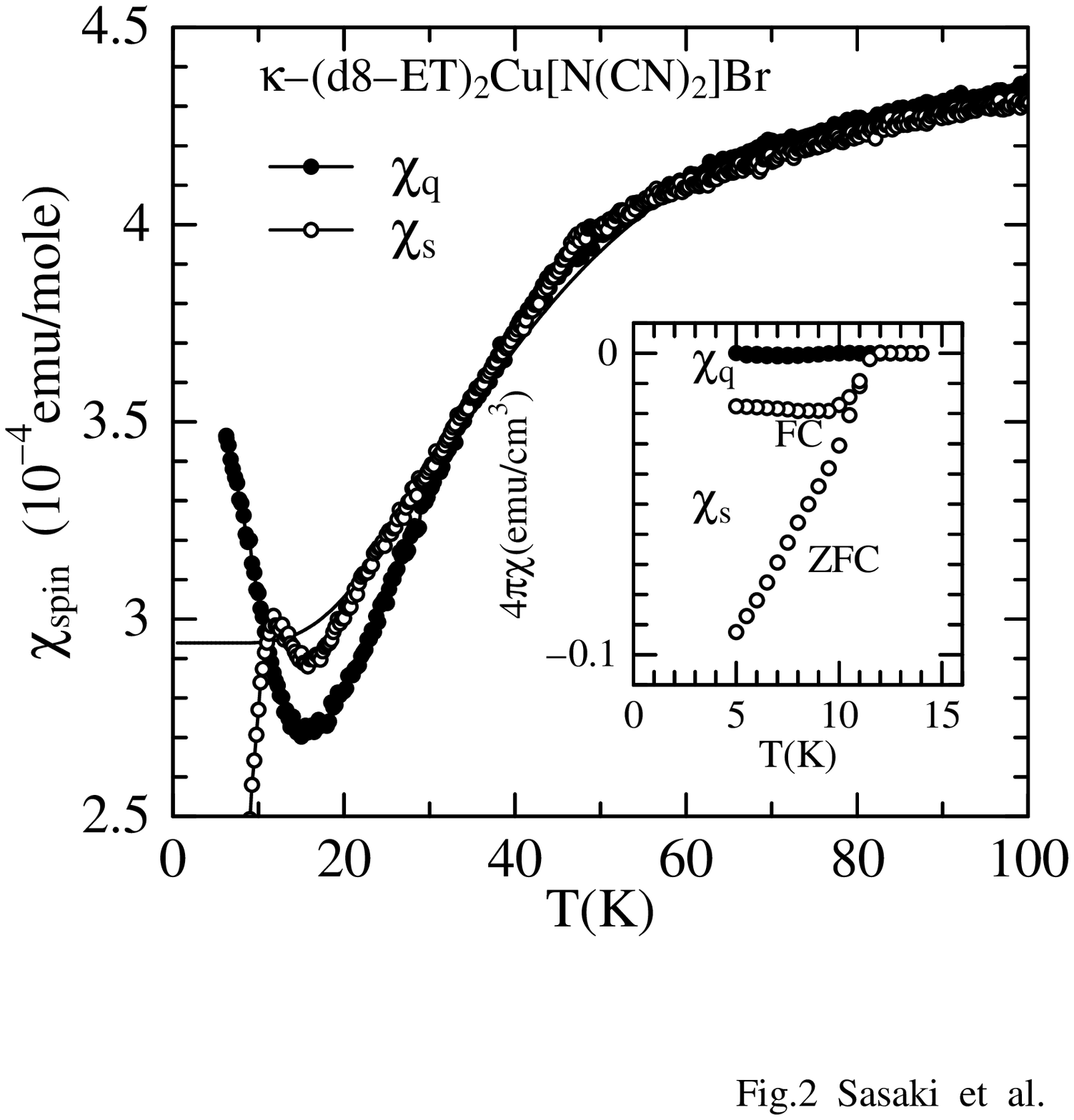}
\caption{Temperature dependence of the magnetic susceptibility of $\kappa$-(d8-ET)$_{2}$Cu[N(CN)$_{2}$]Br in 5~T after the slow cooled ($\chi_{\rm s}$) and quenched ($\chi_{\rm q}$) processes. Inset demonstrates the superconducting transition in $\chi_{\rm s}$ and no transition in $\chi_{\rm q}$ in 0.5~mT.}
\end{figure}

\begin{figure}
\includegraphics[viewport=2cm 7cm 20cm 24cm,clip,width=0.9\linewidth]{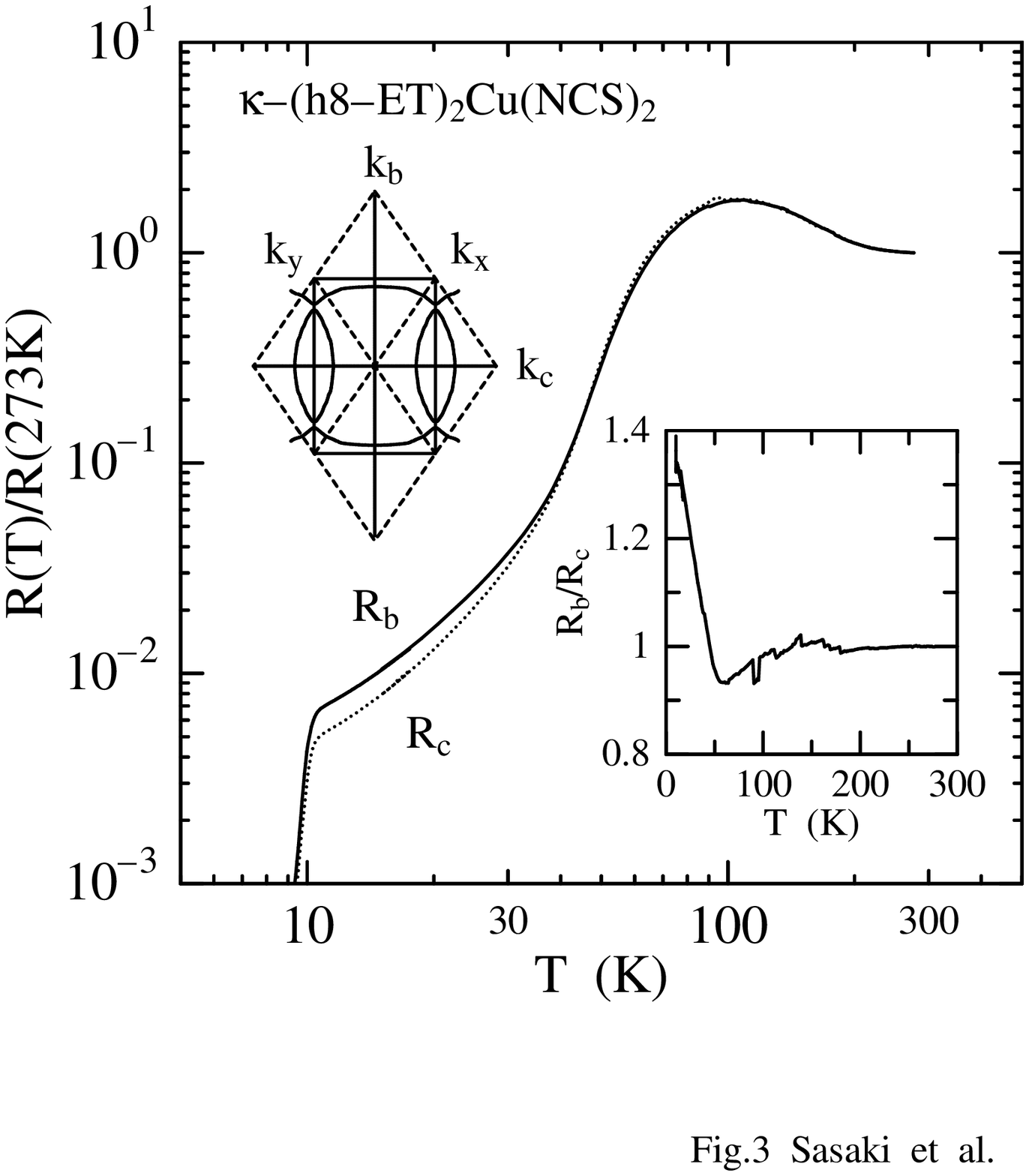}
\caption{Temperature dependence of the normalized resistance along the $b$ and $c$-axes of $\kappa$-(h8-ET)$_{2}$Cu(NCS)$_{2}$.  Top left inset shows the Fermi surface and the Brillouin zone (the solid rectangular with the $k_{b}$ and $k_{c}$-axes).  The dotted diamond with the $k_{x}$ and $k_{y}$-axes is the extended magnetic Brillouin zone in the similar coordinate style of the high-$T_{c}$ cuprates.}
\end{figure}

\begin{figure}
\includegraphics[viewport=2cm 7cm 20cm 24cm,clip,width=0.9\linewidth]{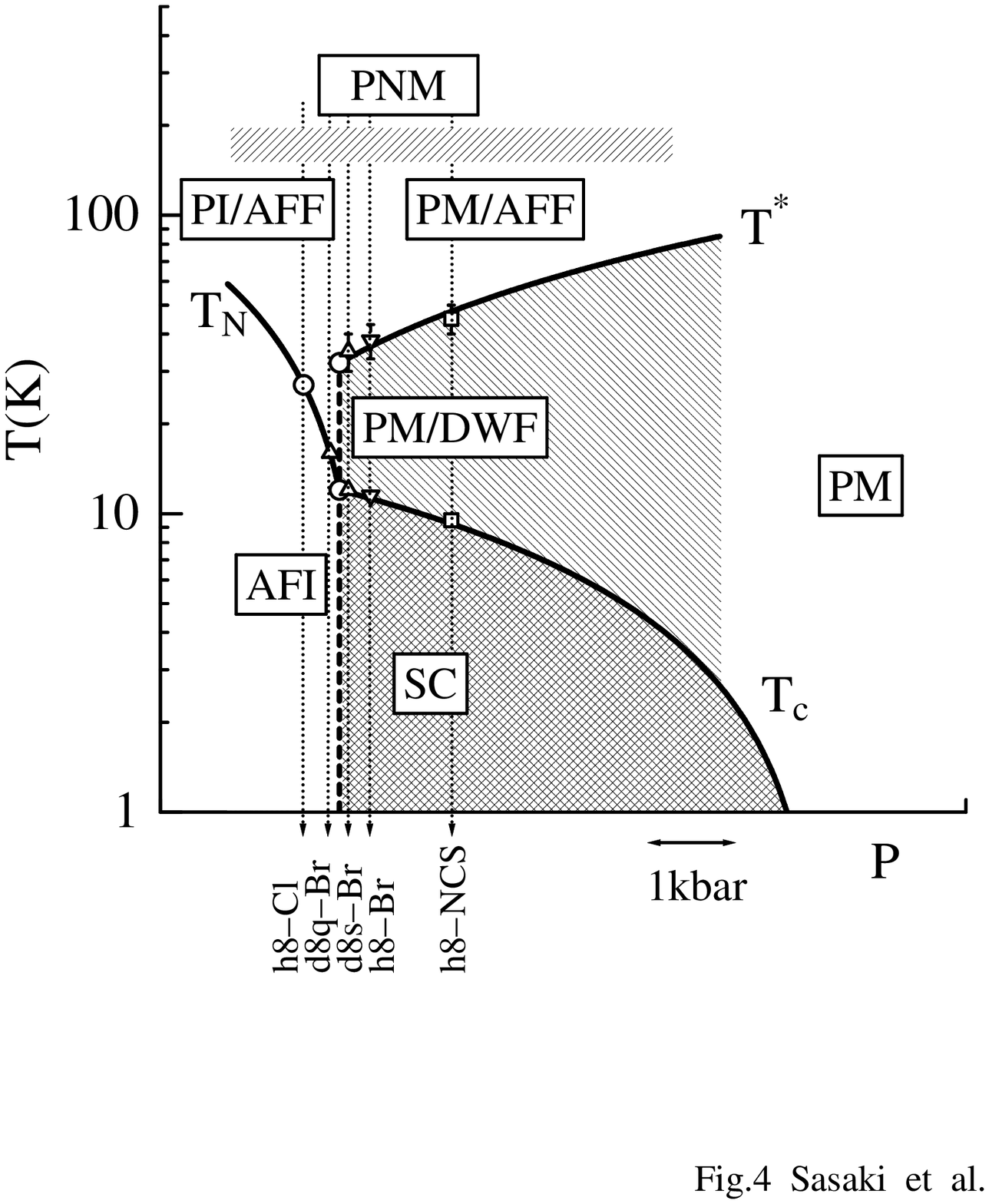}
\caption{Phase diagram of $\kappa$-(ET)$_{2}$$X$. The data of $\kappa$-(h8-ET)$_{2}$Cu[N(CN)$_{2}$]Cl and the pressure dependence of $T_{c}$, $T_{\rm N}$ and $T^{*}$ refer to the pressure studies in Refs. 12, 28--30.  Solid and dashed lines indicate the second and the first-order transitions, respectively.}
\end{figure}


\end{document}